\documentstyle[12pt]{article}
\begin{document}
\baselineskip .3in
\begin{titlepage}
\begin{center}{\large{\bf On some properties of Neutrino in the Early Universe }}
\vskip .2in S.Mani, A.Sagari, B.Chakrabarti$^{+}$  and
A.Bhattacharya $^{\ddag}$
\end{center}
\vskip .1in
\begin{center}
Department of Physics, Jadavpur University \\
Calcutta 700032, India.\\
\end{center}

\vskip .3in {\centerline{\bf Abstract}} The properties of the
neutrino in the early universe have been investigated
incorporating a small inhomogeneity in the mass density of the
early universe. Dependence on this factor is found in studying
mean free path and mass bound of neutrinos. The oscillations of
neutrinos flavours have been studied by assuming a free wave
packet to represent the time progression of the neutrino yielding
interesting results.
\vskip .5in
 {$^{\ddag}$e-mail: pampa@phys.jdvu.ac.in}
\vskip .2in
  {$+$e- mail: ballari$\_$chakrabarti@yahoo.co.in

  Permanent Address: Department of Physics, Jogamaya Devi
  College, Calcutta, India.}
 \end{titlepage}

\newpage
{\bf Introduction}

The neutrino was the first elementary particle proposed that was
not a constituent of ordinary matter, and even now the true nature
of the particle is not clearly understood. Neutrinos, we find, are
involved only in weak interactions and the Standard Model (SM)
predicts them to be mass less. A large body of observational and
experimental efforts has been devoted to verifying whether
neutrinos are massive and oscillate from one flavour to another
[1-8]. The Super Kamiokande experiments [1] suggested electron
neutrino mass $<$ 0.07eV. The experiment also found evidence for
neutrino oscillations and hence mass in atmospheric neutrinos.
Current limits on neutrino mass suggest that $m_{\nu_{e}}< 2 eV$,
$m_{\nu_{\mu}}< 0.19 MeV$ and $m_{\nu_{\tau}} <18.2 MeV$ [9].
Dodelson et al have derived an upper limit for the sum of neutrino
masses as $ \sum m_{\nu}<0.17eV $ ($95percentage$ CL) [10]. Though
experimental estimates of neutrino mass are far from being
perfect, no evidence for neutrinos being mass less is available as
yet [11]. The expanding universe puts a serious constraint on the
neutrino mass. The helium abundance problem does not have any
satisfactory explanation in the theory of thermonuclear fusion or
in nucleosynthesis processes in stars, but may be explained by
primeval nucleosynthesis. At cosmic temperatures greater than
$10^{13}K$, there were few neutrons and photons and the universe
had a large number of electrons, positrons and neutrinos. The
light elements (deuteron) gradually synthesized to helium
abundance and to relative abundance of baryon to photon or
neutrinos. The upper bound on the primordial helium provides an
upper bound on the total energy density $\rho$ at that moment of
the nucleosynthesis which is expressed in terms of the allowed
number of light neutrino families as $N_{\nu}<3+ \delta N_{\nu}$
[12] where $\delta N_{\nu}$ includes any extra contribution beyond
three left- handed neutrinos. The neutrino has a significant
contribution to the cooling of the universe. Hot systems radiate
their excess energy usually as photons or neutrinos. There are
various processes by which the neutrinos were produced in the
early universe such as neutrino pair bremsstrahlung by nucleons,
collective electron plasma excitation, etc. To the first order of
approximation, the relative order of various processes depends on
the equations of state. The neutrinos are supposed to be mass less
in the SM, but Gerstein [13] proposed the massive neutrino from
observational cosmology and relic neutrino abundance. In the SM
the light neutrinos are kept in equilibrium with the charged
lepton and photon in the cosmic plasma due to weak interactions.
But above a temperature of T=1 MeV, as the expansion rate of the
universe increases rapidly, it leads to an early freeze out of the
weak interaction and the number of neutrinos becomes a constant.
The number of photons increases by $e^{+} e^{-}$ annihilation. It
has been suggested that the present number density of neutrinos
per species is $n_{\nu_{i}}=3/11 n_{\gamma} = 102/cm³$ [12]. Hence
the Big Bang cosmology which establishes the expansion of the
universe from extremely high temperature and density and Hubble
expansion has direct consequences on the neutrino density in the
universe. Theoretically the SM can include the possibility of
massive neutrinos. As the neutrino mass eigen states and weak
eigen states are not necessarily coincident, there may exist
neutrino mixing as described by the Cabbibo angle in the GIM
scheme [14] and generalized in Kobayashi- Maskawa [15] six- quark
scheme. Neutrino oscillation is the most reliable observation that
predicts neutrino mass. It is a phenomenon in which a neutrino of
a particular leptonic flavour, over time, transforms to a
different flavour. The idea of neutrino oscillation was first
introduced by Bruno Pontecorvo [16] in 1957 and subsequently the
solar neutrino deficit was observed. Solar neutrino deficit could
be explained theoretically by a vacuum oscillation of neutrino
flavours. A number of experiments followed by Davis [17], Hirata
[18] and others. The experiments to understand and observe
neutrino oscillations have been many and multifarious in approach
[1-3]. The MINOS experiment in 2006 [2] involved, sending a high
intensity beam of muon neutrinos to a particle detector, where it
was found that a significant fraction of the original particles
had disappeared. The NoVA experiment [3], designed and developed
by Fermi lab, is going to send neutrino beams through distances
$\sim$ 800 km to obtain improved measurements of mass differences
among the neutrino flavours. It also hopes to shed light on the
ordering of neutrino mass states- the data is expected to be
available by 2010-11. The current limits on oscillation parameter
are $ \Delta m_{21}^{2}=\Delta
m_{sol}^{2}=8.0^{+0.6}_{-0.4}\times10^{-5} eV^{2}$ [4] and $\Delta
m_{31}^{2}=\Delta m_{32}^{2}=\Delta
m_{atm}^{2}=2.4^{+0.6}_{-0.5}\times10^{-5} eV^{2}$ [5].

We have studied the evolution of the universe with a mass fractal
dimension 'd' as an exponent of the density [19] and come across
many interesting consequences for the early universe lying between
the matter and radiation dominated eras. The speed of sound in the
mixed phase of radiation and matter dominated era has already been
investigated by us [20] considering the cosmological constant to
be zero. In the present work, we have studied the same for a non-
zero value of the cosmological constant. We have investigated
neutrino opacity in the early universe taking into account the
mass fractal dimension 'd' which has been incorporated in the FRW
model of the universe. In inflationary cosmology, the notion of
fractal dimension appears at the stage where at each particular
point it is possible to consider the universe as a Friedmann
universe described by a single scale factor. We use the fractal
concept to study the formation and structure of the universe
between matter and radiation dominated eras and explore the
evolution of the early universe. We have investigated neutrino
opacity in the early universe with this approach. The equation of
state for the universe is found to yield some interesting results
on neutrino density, mean free path of neutrinos and a mass bound.
Vacuum oscillations of neutrino flavors have also been studied
considering the evolution of the mass eigen states as free wave
packets. The neutrino states are wave packets with a certain
spreading in momentum, and hence such an approach is justified. An
expression for probability of oscillation (considering only two
flavors) is deduced and compared to other similar results.

\vskip .3in {\bf Velocity of sound}

 The standard cosmological
model of the universe is based on FRW metric. The Einstein's
equation runs as [21],
\begin{equation}
\lbrack{\frac{\rm dR(t)/dt}{\rm R}}\rbrack^2 = {\frac{\rm 8\pi G
\rho}{\rm 3}} - \frac{\rm K}{\rm R^{2}(t)} + \Lambda
\end{equation}
where the symbols have their usual meanings. The energy
conservation demands that:
\begin{equation}
\frac{\rm d( \rho R^{3})}{\rm dR} = -3PR^{2}
\end{equation}
where P is the isotropic pressure. With a small inhomogeneity in
the universe through 0 $<$ d $<$ 1 [22] where energy behaves as,
$\rho$(R) $\sim$ $R^{-d-3}$ we have,
\begin{equation}
\frac {\rm d(R^{-d})}{\rm dR} = -3PR^{2}
\end{equation}
We presume an adiabatic expansion of state as P = A$\rho^{\gamma}$
($\gamma$ $>1$)  corresponding to the rapid expansion era of the
early universe. The energy conservation equation can be
generalized through a fractional differentiation of nth order as
[22],

\begin{equation}
\frac{\rm d^{n}R^{-d}}{\rm dR^{n}} = -3AR^{-(d+3)\gamma+2}
\end{equation}
where n is, in general, a fractional number. As density
fluctuation scales like a fractal as $R^{-d-3}$, we have replaced
the flux $\frac {\rm d(R^{-d})}{\rm dR}$ in (3) by $ \frac{\rm
d^{\mu}R^{-d}}{\rm dR^{\mu}}$. With $\mu$ = 1 , $\gamma$ = 1, we
recover the usual conservation equation of (3). For a flat
universe (k =0) we get from (1) with $\Lambda$ = 0 and $\rho$
$\sim$ $R^{-d-3}$ [22],
\begin{equation}
R(t) \sim t^\beta
\end{equation}
where $\beta$ = 2/d+3. 0$<$ d $<$ 1 corresponds to the coexistence
phase of matter and radiation dominated universe and d = 0
represents the matter dominated era whereas d = 1 corresponds to
the radiation dominated era. We have also derived the result $t=
\beta H^{-1}$ where H is the Hubble parameter [19]. With the non
zero value of the cosmological constant the energy conservation
equation runs as [23],

\begin{equation}
\frac{\rm d(\rho R^{3})}{\rm dR} + P\frac{\rm dR^{3}}{\rm dR} +
\frac{\rm \alpha R^{3}}{\rm 3}.\frac{\rm d \Lambda }{\rm dR} = 0
\end{equation}

Assuming the cosmological constant '$\Lambda$ ' as an explicitly
scale dependent quantity as,
\begin{equation}
\Lambda \sim \beta^{\prime}R^K
\end{equation}
where $\beta^{\prime}$ is an arbitrary constant and
$\beta^{\prime}$ $>$ 0. We may recast the equation (6) as:
\begin{equation}
\frac{\rm d(\rho R^{3})}{\rm dR} + \frac{\rm \alpha K
\beta^{\prime} }{\rm 3 (K+3)}.\frac{\rm d R^{K+3} }{\rm dR} =
-3PR^2
\end{equation}
where $\alpha$ = $\frac{3}{8} \pi G$. With adiabatic pressure P =
A $\rho^\gamma$ ($\gamma$ $\rightarrow$ 1 ), $\rho$ $\sim$ $R^K$,
where $K=-d-3$ (the energy conservation equation (6) is satisfied
provided '$\Lambda$' scales in the same manner as the density
distribution scales in the early universe) we have,
\begin{equation}
-3A = K (\frac{\rm \alpha\beta^{\prime}}{\rm 3}+1) + 3
\end{equation}
For radiation and matter dominated era where A $>$ 0, we must have
$K(\frac{\rm \alpha\beta^{\prime}}{\rm 3}+1)$ + 3 $<$ 0. Hence,
\begin{equation}
\mid K\mid > 3/ (\frac{\rm \alpha\beta^{\prime}}{\rm 3}+1)
\end{equation}
Hence the velocity of sound can be expressed as ,
\begin{equation}
v_{s} ={( \frac{\rm \delta P}{\rm \delta \rho})}^{\frac{\rm1}{\rm
2}}
\end{equation}
Thus we have for $\Lambda$ = 0 and $\gamma$ $\rightarrow$ 1 [20],
\begin{equation}
v_{s} \propto d^{ \frac{\rm 1}{\rm 2}}
\end{equation}
and  for $\Lambda$ $\sim$ $\beta^{\prime}R^{K}$,
\begin{equation}
v_{s} \simeq \frac{\rm1}{\rm \sqrt3} \lbrack \frac{\rm
(3+\alpha\beta^{\prime})d}{\rm3}+\alpha\beta^{\prime}\rbrack
^{\frac{\rm1}{\rm2}}
\end{equation}

For $\Lambda$ = 0 and $\gamma$ $\rightarrow$ 1, this reduces to
$v_{s}\sim \surd d$. The velocity of sound, we see, depends on the
mass fractality  or the inhomogeneity of the medium.

{\bf Mean free path}

 \vskip .3in
 In the early universe during the primeval nucleosynthesis process the neutrinos produced
 lose their energy due to scattering from the electron and the nucleon at the energy scale
 T $ \approx$ $10^{15}$K. A neutrino propagating in a medium of electron gas has the mean free
 path [24],
\begin{equation}
\lambda_{\nu} \approx \frac{\rho^{-4/3}}{\varepsilon^{3}_{\nu}}
\approx \frac{R^{1.33(d+3)}}{\varepsilon^{3}_{\nu}}
\end{equation}
where $\lambda_{\nu}$ is the neutrino mean free path, $\rho$ is
nuclear density and $\varepsilon_{\nu}$ is neutrino energy. For a
degenerate Fermi gas of neutrons we arrive at a neutrino mean free
path,
\begin{equation}
\lambda_{\nu} = \frac { \rm \kappa_{T}}{ \rm
\epsilon^{2}_{\nu}(kT)n^{2}_{n}}
\end{equation}
 Where $\kappa_{T}$ is the bulk modulus of the medium, $\epsilon^{2}_{\nu}$ is the neutrino energy
 and assumed to be $\epsilon_{\nu}$ =kT and $n_{n}$ is neutrons/$fm^{3}$.

With $v_{s}$ = $\sqrt{ \frac{\gamma P}{\rho}}$ (where $\gamma$ is
adiabatic constant of the medium) , we can rewrite the above
expression as:

\begin{equation}
\lambda_{\nu} \approx \frac {\rm v_{s}^2\rho}{\gamma(kT)^3n_{n}^2}
\end{equation}
with R = $t^{\frac {\rm 2}{\rm {d+3}}}$, $t=\beta H^{-1}$ [19], we
arrive at,
\begin{equation}
\lambda_{\nu} \sim(d+3)^{2}H^{2}
\end{equation}

Similarly taking into the contribution from $\Lambda$, we get,
\begin{equation}
\lambda_{\nu} \sim(d+3)^{2}H^{2}
\end{equation}

Thus we find from the equation (18) and (19) that the mean free
path of neutrino depends on the mass fractal dimension of the
medium. The expansion of the universe affects the mean free path
of the neutrino and consequently affects the scattering cross
section.

{\bf Mass Bound}

One of the most important cosmological constraints on stable
neutrino is its mass bound. In SM the interaction is kept in
equilibrium with charged lepton and photons until a temperature T
$\approx$ 1 Mev. In the early universe during the nucleosynthesis
process the numbers of photons and neutrinos are the same at about
T $\simeq$ $10^{15}$K [12]. After that the rapid expansion of the
universe freezes the weak process making the neutrino density
fixed. However the $e^+$$e^-$ annihilation increases the photon
density. In the context of investigating the fractal structure of
the universe [22], we have come to the expression,

\begin{equation}
\frac{ \rm 2\pi}{\rm3}(d+3)^2 G\rho t^2 = 1
\end{equation}

 where 'd' is the  mass fractal dimension and $\rho$ = (g/2)$\rho_\gamma$ where $\rho_\gamma$
 is the photon energy density which may be considered to be equal to the neutrino density $\rho_\nu$
 in the early phase of the universe between matter and radiation dominated era,
 g is the corresponding degrees of freedom. Now from the above equation we have estimated the neutrino energy density
 with $\rho$$\approx$(g/2)$\rho_\nu$, t= $10^{-4}$sec, G=6.673$\times10^{-8}$ $cm^3$
 $gm^{-1}sec^{-2}$,as
\begin{equation}
\rho_{\nu}= \frac{\rm0.1413}{\rm g(d+3)^2}\times10^{16} gm/cm^{3}
\end{equation}
Now the number density of the neutrino $n_{\nu}$ is almost equal
to the number of photons at the temperature of the order of
T=$10^{15}K$ and $n_{\gamma}$ = $n_\nu$ $\approx$ $T^3$ [25].
 If the neutrinos be massive, the energy density of neutrino
is $\rho_\nu$ = m$_\nu$$n_\nu$ where m$_\nu$ is the mass of the
neutrino. Thus from the expression (21) we have,
\begin{equation}
m_{\nu} = \frac{\rm 1.43}{\rm g(d+3)^2}\times10^{-30}gms
\end{equation}
To get an estimate of the mass of the neutrino we use g$\approx$
427/4 at the electroweak scale [26] and come across $m_{\nu}$ =
2.29$\times10^{-32}$/$(d+3)^2$gms.
 Thus the mass of the neutrino depends on the mass fractal dimension of the universe and $m_{\nu}$
 is found to lie between 14eV and 76eV between radiation (d=1) and matter (d=0) dominated era.
 Assuming the left handed neutrino are ever in equilibrium in the primordial plasma, Gerstein et al
 [13] have predicted m$_{\nu_{i}} \leq$ 37 eV in a matter dominated
universe with zero cosmological constant.
 They have emphesized to the fact that the nature of the matter, radiation and cosmological constant
 influence the value of $\Omega_{0}$ very much. With 0.25 $\leq$ $\Omega_{0}$ $\leq $0.4,
 they have obtained 0.27 $<$ $\Sigma$ m$_{\nu}$$_{i}$ $<$ .37 ev  (where i varies from 1 to 3).
 m$_{\nu}$ has also been estimated from $\frac{\rm n_{B}}{\rm n_{\gamma}}$ [27].
 The fraction of the critical mass taken by  neutrino has been estimated to be
 $\approx$ 20 eV assuming $\Omega$ = 1 (which implies the neutrino provide enough mass to close the universe).
 Here we observe that inhomogeneous nature or mass
fractal dimension limits the mass of the neutrino in between 14eV
to and 76eV in the coexistence phase of matter and radiation. So
it may be suggested that if we consider the universe starts with a
small inhomogeneity, the energy density or the mass is greatly
influenced by it.

{\bf Neutrino Oscillation}

Neutrino oscillations, till date, remain the most promising avenue
of exploration to find neutrino mass. We study the vacuum
oscillation of neutrinos considering their evolution as free wave
packets. We observe that significant suppression in one type of
flavour demands very small momentum difference in the
corresponding flavours. Theoretically neutrino mixing gives rise
to the phenomenon of neutrino oscillation. In this phenomenon a
neotrino born in an eigen state, after a certain time interval,
finds itself in a mixture of states. Confining ourselves to the
first two generations, we assume that the flavour eigen states
$\nu_{e}, \nu_{\mu}$ are produced at time t=0. The corresponding
mass eigen states are $\nu_{1}, \nu_{2}$. We express the weak
eigen states as a linear combination of the mass eigen states [27]
with $\theta$ as an angle that parametrizes the mixing, which can
be calculated if we know the interaction that gave rise to the
masses. A non zero value of $\theta$ implies that some neutrino
masses are non- zero and that the mass eigen states are non-
degenerate. If $m_{\nu_{i}}$=0, then there is no way to
distingiush the weak eigen states from the mass eigen states, so
the states could always be expanded in a new set and the angle
$\theta$ rotated to zero. For the time evolution of the mass eigen
states we consider the propagation of the free particle as
suggested by Kleber [28]. The expression runs as

\begin{equation}
\nu_{i}(r,t)=\frac{m_{i}}{2iht\pi}^{3/2} e^{i m_{i}r^{2}/h t} \int
d^{3}r^{\prime}e^{im_{i}(r^{\prime})^{2}/2ht}
e^{-im_{i}rr^{\prime}/ht}\nu_{i}(r,0)
\end{equation}

 Without any loss of generality we assume that $\nu_{i}(0)$ is
 concentrated around $\overrightarrow{r}=0$. Hence we may rewrite
 the the previous expression as

\begin{equation}
\nu_{i}(r,t)=\nu_{i}(0,0)\frac{m_{i}}{2iht\pi}^{3/2} e^{i
m_{i}r^{2}/ht} \int^{a}_{0} 4\pi (r^{\prime})^{2} dr^{\prime}
e^{im_{i}(r^{\prime})^{2}/2ht} e^{-im_{i}rr^{\prime}/ht}
\end{equation}

wherte 'a' is the cut- off parameter which is presumed to
represent the oscillation length. The expression represents the
evolution of the mass eigen state at some time t ($t>0$). At time
t=0 we assume the mass eigen state and weak eigen state to be
identical, and $\nu_{\mu}$(0) is orthogonal to $\nu_{e}$(0). The
mass eigen states vary with time as per the above equation, and
they are free particles after they are produced. Expressing the
weak eigen states as a combimation of the mass eigen states we see
the time evolution through the wave packet description. On writing
$\nu_{\mu}$(t) we find that at time t=o what was a pure state
$\nu_{\mu}$(0) has now become a mixed state containing
contribution from both $\nu_{\mu}$(0) and $\nu_{e}$(0). The
probability that the $\nu_{\mu}$ beam after a time t will contain
$\nu_{e}$, on calculation is found to be given by,
\begin{equation}
p(\nu_{\mu} \rightarrow \nu_{e}) =|<\nu_{e}(0)|\nu_{\mu}(t)>|^{2}
=\frac{Bsin^{2}2\theta}{t}[1-cos(E_{2}-E_{1})t]
\end{equation}

where $B= \frac{\pi m a^{2}}{2i^{3}\hbar}$ and $E_{i}$ are the
energies corresponding to the mass eigen states $\nu_{i}$. This
describes the simplest type of neutrino flavour oscillation, in
which the amplitude of oscillation is found to depend on the
mixing angle $\theta$ and time t. The oscillation frequency is
found to be proportional to $\Delta m^{2}$ and momentum p, by
rewriting $(E_{2}-E_{1})t \simeq \frac{m_{2}^{2}-m_{1}^{2}}{2p}t$.
We define the oscillation length 'a' by putting
$cos(\frac{m_{2}^{2}-m_{1}^{2}}{2p})t \simeq cos \frac{2\pi
x}{a}t$ where $x \simeq ct$ is the distance travelled by the beam
in time t. Now, $a=\frac{4 \pi p}{m_{2}^{2}-m_{1}^{2}}$, where 'a'
is an effective length that determines the distance over which one
might expect to see the effect.

\vskip .3in

{\bf Conclusion}

 In the present work we have investigated some
interesting feature of neutrino properties in the early universe
in between matter and radiation dominated era. The velocity of
sound when calculated fora non- zero value of the cosmological
constant, is found to depend on the mass fractal dimension 'd'. We
find that the velocity of sound in radiation dominated phase is
much greater than in the matter dominated phase, as expected.

An expression for the mean free path of the neutrino is deduced
and hence we find that neutrino opacity is influenced by the
inhomogeneity of the medium, which we have incorporated through a
mass fractal dimension, in the early universe. It may be pointed
out here that although the nucleon-neutrino scattering process is
not that much dominant in the early universe, it would be very
interesting to investigate the influence of the fractal dimension
in the scattering process via nucleon-neutrino process during the
primeval nucleosysnthesis. The mass of the neutrino has been found
to be influenced by the fractal dimension in the present work,
whereas Gerstein et al [13] has emphasized the fact that the mass
of the neutrino is greatly influenced by the value of
$\Omega_{0}$. It would be relevant to mention here that the large
neutrino mass could make the universe finite and close. The
critical density has the value $\approx$ $58 \times10^{-30}$
$gm/cm^{3}$. Considering present day photon and neutrino density
almost equal to (400/cm$^{3}$)$^{7)}$, the contribution from the
neutrino mass will exceed the critical density if $m_{\nu}$ is
more than 10$^{-32}$ gms. The last word on neutrino mass, is
however far from being spoken, especially since we are still in
the dark regarding the nature of dark matter.

The mass bound of the neutrino as we estimate it is between 14- 76
eV. Almost all recent research agrees on putting $\nu_{e}$ mass at
les than 50 eV, with even lower masses also being suggested [1,9].
The sum of the three neutrino masses is required to be very small
in order that the universe does not become a close one.

Neutrino oscillations have been studied in detail with the results
throwing up a time- dependent amplitude of oscillation. Also the
frequency of oscillation provides us with a relation connecting
oscillation length, the neutrino momentum and mass difference
squared. Taking the mass difference squared as $7.59\times 10^{-5}
eV^{2}$ and beam energy 2.754MeV from SNO results [29] we
calculate the oscillation length to be 86 km. For the same mass
difference if the beam energy is 7 MeV [1], the oscillation length
is found to be 220 km. It may be mentioned that in the Super
Kamiokande experiment the detector is placed at a distance of
approximately 250 km to observe neutrino oscillations. As per
current estimates of neutrino masses, the mass difference squared
is expected to be very small, which can only be detected for large
value of oscillation lengths.

\pagebreak

{\bf References}

\noindent [1]. Y Fukuda et al; Phys. Rev. Lett. {\bf 81}, (1998),
1158; Phys. Rev. Lett. {\bf 81}, (1998), 1562.

\noindent [2]. MINOS Collab; Phys. Rev. Lett. {\bf 97}, (2006),
191801.

\noindent[3]. R Plunkett et al; J. Phys.: Conf. Ser. {\bf 120},
052044, (2008).

\noindent [4]. KamLAND Collab; arXiv: 0801.4589.

 \noindent [5]. LSND Collab; Phys. Rev. {\bf D 64}, (2001), 112007.

\noindent [6]. K2K Collab; Phys. Rev. {\bf D 74}, (2006), 072003

\noindent [7]. MiniBooNE Collab;Phys. Rev. Lett. {\bf 99}, (2007),
231801.

\noindent [8]. SNO Collab; Phys. Rev. {\bf C 72}, (2005), 055502.

\noindent [9]. PDG; Phys. Lett. {\bf B 667} Issues 1-5, (2008),
Pg.517-519.

\noindent [10]. S Dodelson et al; New Astronomy Review; {\bf 50},
(2006), 1020.

\noindent [11]. K V L Sharma; Int. J. Mod. Phys. {\bf A 10},
(1995), 767; P H Bucksbaum, Weak Interactions of Leptons and
Quarks, Cambridge University Press (Cambridge), (1983), Pg. 369.

\noindent [12]. G Gelmini et al; Rep. Prog Phys. {\bf 58}, (1995),
1207.

\noindent [13]. S Gerstein,Ya. Zeldovich., Zh. Eksp. Theor. Fiz
Pis {\bf 4}, (1972), 174.

\noindent [14]. S L Glashow et al; Phys. Rev. {\bf D 2}, (1970),
1285.

\noindent [15]. M Kobayashi, T Maskawa; Prog. Theor. Phys. {\bf
49}, (1973), 652.

\noindent [16]. B Pontecorvo, Gribov; (neutrino astronomy and
lepton charge)

\noindent [17]. R Davis et al; Proc. 21st Intl; Cosmic Ray Conf.
edited by R J Protheroc, Pg. 143

\noindent [18]. K Hirata et al; Phys. Rev. Lett. {\bf 66}, (1991),
9; Phys Rev {\bf D 44}, (1991), 2241.

\noindent [19]. S N Banerjee et al; Mod. Phys. Lett. {\bf A 12},
(1997), 573; B Chakrabarti et al; Astroparticle Phys. {\bf 19},
(2003), 295; S N Banerjee et al; Astroparticle Phys. {\bf 12},
(1999), 115.

\noindent [20]. A Bhattacharya et al; Mod. Phys. Lett. {\bf A 14}
No 15, (1999), 951.

\noindent [21]. S Weinberg; Gravitation and Cosmology, John Wiley
and Sons, (1972), 472.

\noindent [22]. A Bhattacharya et al.; Phys. Lett. {\bf B 492},
(2000), 233.

\noindent [23]. A.M. Abdel Rahman; Phys. Rev. {\bf D 45}, (1992),
3497.

\noindent [24]. J.M. Irvine; Neutron Star, Oxford University
Press, (1978), Pg 43.

\noindent [25]. F.L. Zhi and L.S.Xian; Creation of the Universe,
World Scientific, (1982), 92.

\noindent [26]. S.Sarkar; Rep. Prog. Phys. {\bf 59},(1996), 1493.

\noindent[27]. Gordon Kane, Modern Elementary Particle Physics,
Addision Wesley Publishing Company, (1987), chapter 29.

\noindent [28]. M Kleber; Phys. Rep.{\bf 236}, (1994), 331.

\noindent [29]. B Aharmin et al; Phys. Rev. Lett. {\bf 101},
(2008), 111301.

\end{document}